
\input phyzzx

\baselineskip=22pt
\overfullrule=0pt
\font\twelvebf=cmbx12
\nopagenumbers
\footline={\ifnum\pageno>1\hfil\folio\hfil\else\hfil\fi}
\line{\hfil IP-ASTP-06-95}
\line{\hfil CU-TP-679}
\line{\hfil hep-th/9503200}

\vglue .4in
\centerline {\twelvebf  Quantum Nucleation of Vortex String Loops}

\vskip .5in
\centerline{\it Hsien-chung Kao$^\dagger$ and Kimyeong Lee$^*$  }

\vskip .1in
\centerline {$^\dagger$Institute of Physics, Academia Sinica}
\centerline {Taipei, Taiwan}
\centerline {and}
\centerline {$^*$Physics Department, Columbia University}
\centerline {New York, New York 10027}
\vskip .2in

\baselineskip=20pt
\overfullrule=0pt

\vskip .5in
\centerline {\bf Abstract}
\vskip .1in

We investigate quantum nucleation of vortex string loops in the
relativistic quantum field theory of a complex scalar field by using
the Euclidean path integral.  Our initial metastable homogeneous field
configurations carry a spacelike current. The path integral is
dominated by the $O(3)$ symmetric bounce solution. The nucleation rate
and the critical vortex loop size are obtained approximately.
Gradually the initial current will be reduced to zero as the induced
current inside vortex loops is opposite to the initial current. We
also discuss a similar process in Maxwell-Higgs systems and possible
physical implications.

\vfill

\footnote{}{$^*$ Email Address: klee$@$cuphyf.phys.columbia.edu}
\footnote{}{$^\dagger$ Email Address: hck$@$phys.sinica.edu.tw}
\vfill\eject

\def\pr#1#2#3{Phys. Rev. {\bf D #1}, #2 (19#3)}
\def\prl#1#2#3{Phys. Rev. Lett. {\bf #1}, #2 (19#3)}

\def\np#1#2#3{Nucl. Phys. {\bf B #1}, #2 (19#3)}
\def\pl#1#2#3{Phys. Lett. {\bf B #1}, #2 (19#3)}
\def\ibid#1#2#3{{\it ibid}. {\bf #1}, #2 (19#3)}

\def\vx{{\bf x}}
\def\vk{{\bf k}}

\REF\rSchwinger{ J. Schwinger, \pr{82}{664}{51}.}

\REF\rAffleck{ I.K. Affleck and N.S. Manton, \np{194}{38}{82};
I.K. Affleck, O. Alvarez, and N.S. Manton, \ibid{B 197}{509}{82}.}

\REF\rGiddings{D. Garfinkle, G. Horowitz, and
A. Strominger, \pr{43}{3140}{91}; D. Garfinkle, S.B. Giddings and A.
Strominger, \ibid{D 49}{958}{94}; F. Dowker, J.P. Gauntlett, S.B.
Giddings, G.T. Horowitz, \ibid{D 47}{3265}{94}; P. Yi,
\pr{51}{2813}{95}.}

\REF\rBachas{C. Bachas and M. Porrati, \pl{296}{77}{92}.}

\REF\rGuth{ R. Basu, A.H. Guth and A. Vilenkin, \pr{44}{340}{91};
R. Basu and A. Vilenkin, \ibid{D 46}{2345}{92}; J. Garriga,
\ibid{D 49}{6327}{94}; M. Axenides and A.L. Larsen, ``Charged Cosmic
String Nucleation in deSitter Space'', Nordita 94-38-P,
hep-th/9409005 (1994).  }

\REF\rDonnelly{ J.S. Langer and M.E. Fisher, \prl{19}{560}{67};
R.J. Donnelly, ``Quantized Vortices in Helium II,''
Cambridge University Press, Cambridge (1991); D.R.  Tilley
 and J. Tilley, ``Superfluidity and Superconductivity,'' 3rd ed.,
Adam Hilger, Bristol (1990).}

\REF\rAo{ P. Ao and D.J. Thouless, \prl{72}{132}{94}; M.J. Stephen,
\prl{72}{1534}{94}.}

\REF\rDavis{R.L. Davis, Physica {\bf B 178}, 76 (1992).}

\REF\rLanger{ J.S. Langer, Ann. of Phys. {\bf 41}, 108, (1967);
S. Coleman \pr{15}{2929}{77}; C.G. Callan and S. Coleman, \pr{16}{1762}{77}.}

\REF\rKL{K. Lee, \pr{50}{5333}{94}.}

\REF\rDavis{ R.L. Davis and E.P.S. Shellard, \prl{63}{2021}{89}; K.
Lee, \pr{48}{2493}{93}. }

\REF\rDuan{J.-M. Duan, Phys.Rev. {\bf B 48}, 4860 (1993);
\prl{72}{586}{94},}

We consider the theory of a complex scalar field with a global $U(1)$
symmetry.  When a field configuration is homogeneous, its conserved
four-current can be timelike, spacelike or lightlike.  Here, we
consider quantum dynamics of classically stable configurations
carrying a spacelike current. We argue that they are quantum
mechanically unstable, and decay through quantum nucleation of vortex
string loops whose current reduces the existing current inside the
ring, leading to the slow decay of the initial current.

When there is a uniform external electric field, electron-positron
pairs could be created via quantum tunneling and will fly apart
because of electrostatic force [1].  In  Yang-Mills
Higgs systems with magnetic monopoles in broken phase,
monopole-antimonopole pairs would be created in a uniform magnetic
field [2].  If the gravity is taken into account, magnetically charged
black hole pairs can be created when there is a uniform external
magnetic field [3].  Similarly, in  open string theories, where
positive and negative charges are attached to the ends of the strings,
open strings would be created by tunneling in a uniform electric field
[4]. In the deSitter spacetime, there can also be a nucleation of
closed vortex strings due to the expanding spacetime [5].

In the dual formulation of the theory of a complex scalar field in
three-dimensional spacetime, vortices appear as charged particles and
the uniform current does as a uniform external electric field.  Thus,
vortex-antivortex pairs will be nucleated and fly apart if there is a
uniform spacelike current. The present work is to understand a similar
process in four-dimensional spacetime. Instead of vortex-antivortex
pairs, closed vortex string loops will be nucleated and expanded. In
condensed matter physics, there is a long history of investigations on
the vortex loop nucleation in superfluids in finite temperature due to
the thermal and quantum effects, whose review can be found in Ref [6].
(Recent discussions about tunneling involving vortices can be found in
Ref.[7].) The global charge current in superfluids is timelike and the
thermal nucleation of vortices in superfluids is due to the relative
velocity between superfluid and normal fluid.

In this paper, we consider the case where the initial current is
spacelike. Clearly such a system cannot be realized naively in
superfluids whose current is timelike.  We, however, hope that this work
enhances our understanding of the vortex string loop nucleation in
general and other related issues like what configuration the system
would settle down to after decay of the initial current. In a similar
context, there has been a recent effort to understand the vortex
string loop nucleation in a scalar field theory[8], whose analysis we
feel is somewhat heuristic as the current the paper assumes  is
time-like.   In addition, we also consider briefly the case
when the scalar field is coupled to the gauge field.   Additional
physical implications will be discussed at the end.

Our treatment of the vortex string nucleation follows the Euclidean
path integral method to calculate the imaginary correction to the
initial energy density [9].  The vortex loop nucleation rate would
then be twice the imaginary part of the false vacuum energy density.
We will consider the leading order contribution and make some comments
about the one-loop correction, which is much smaller than the
uncertainty in our approximation to the leading order contribution.
This work complements the work by one of us in which case the initial
four-current is timelike [10], in which case the physics is completely
different. The quantum nucleation of vortex ring is not possible
energetically. If the initial configuration with time-like current or
a uniform charge density is quantum mechanically unstable, it will
decay by the bubble nucleation and expansion.

This paper is organized as follows: We first discuss the
characteristics of the initial configurations which is classically
stable and spatially homogeneous. Then we discuss the Euclidean path
integral including the vortex strings by using the dual formulation.
We then solve the bounce equation approximately, which allows us to
determine the radius of the string loop at the moment of the
nucleation and the vortex nucleation rate per unit volume.  We then
investigate a similar case in a Maxwell-Higgs system and argue that
the initial configuration with an external current could decay through
the quantum nucleation of vortex string loops. Finally we conclude
with some remarks.

A theory of a complex scalar field $\phi = fe^{i \psi} /\sqrt{2} $
is given by the Lagrangian,
$$ {\cal L } = {1\over 2} (\partial_\mu f)^2 + {1\over 2}
f^2(\partial_\mu \psi)^2 - U(f).
\eqno\eq $$
The conserved current for the global $U(1)$ symmetry is $ J^\mu = f^2
\partial_\mu \psi$ and its conserved charge $Q = \int d^3x J^0 $.
The ground state of the system is given by $<f> = v$, where the global
symmetry is spontaneously broken. In this vacuum, there are  massless
Goldstone bosons and  massive scalar bosons. In the broken phase there
also exist topological global strings around each of which the phase
${\bf \psi}$ changes by $2\pi$.

Here, we are interested in quantum decay of classically stable
homogeneous configurations with a global current $J^\mu$. Depending on
whether the current is timelike or spacelike, we can choose a frame
where there is a uniform charge density but no spatial current or a
uniform spatial current but zero charge density.  In this paper we
consider the case when the current is spacelike. We then choose a
reference frame where only nonzero component of $J^\mu$ is $ J^z = j
$.  For a given current $j$, the  homogeneous $f,\psi$
configuration is given by $ f = u, \psi = -j z /u^2 $. This
homogeneous configuration is static in time as we require the $u$ to
satisfy
$$ j^2 + u^3 U'(u) = 0. \eqno\eq $$

To find out whether such a static homogeneous configuration is stable
classically, let us analyze the field equation under small
fluctuations.  Under small fluctuations around this configuration, $ f
= u + \delta f, \psi = -j z/u^2 + \delta \psi $ and $\delta f ,
\delta \psi \sim e^{i(wt - \vk\cdot\vx)}$, we get a dispersion
relation from the field equation.  The massive mode of mass
$m=\sqrt{j^2/u^4+ U''(u)}$ is always stable but
the massless mode could be unstable.  For small
${\bf k}$, the dispersion relation for the massless mode is
$$ w^2 = k_x^2 + k_y^2 + { u^4 U''(u) - 3j^2 \over
 u^4 U''(u) + j^2} k_z^2.
\eqno\eq $$
This mode is stable only  if
$$  u^4U''(u) - 3j^2 > 0.
\eqno\eq $$
The two conditions, (2) and (4), can be obtained easily by considering
an effective potential $ U_{eff}(f) = -j^2 / 2f^2 + U(f) $. A stable
homogeneous configuration appears as a stable point of this effective
potential.  It is intersting  that the Goldstone mode with the $z$
momentum behaves like a sound wave of speed less than unity.

We now choose a specific potential
$$ U(f) = {\lambda \over 4}(f^2-v^2)^2.
\eqno\eq $$
For this specific potential (5), the condition (2) becomes
$$  j^2  + \lambda u^4(u^2-v^2) = 0.   \eqno\eq $$
When the magnitude of the initial current $j$ is less than a
critical current
$$ j_c = {2\sqrt{\lambda} v^3 \over 3 \sqrt{3}},
\eqno\eq $$
there are two solutions $u_\pm$ of Eq.(6),  $u_- < \sqrt{2/3} v < u_+ <
v $.  The stability condition (4) is satisfied only by the
$u_+ $. The classically stable initial configuration we are
considering here is then given by $f= u_+, \psi = jz/u_+^2$. When the
initial current is very small, Eq.(6) implies that $ u_+ \approx v (1
- j^2/ 2\lambda v^6 ) + {\cal O}(j^4) $. As the initial current gets
larger, $u_+$ decreases to $\sqrt{2/3}v$. There is no solution to
Eq.(6) when $|j| > j_c$.  Thus there is a maximum homogeneous current
the system can endure.

If the initial current is smaller than $j_c$, a classically stable
homogeneous configuration is possible.  However, we know it is always
quantum mechanically unstable by quantum nucleation of vortex string
loops whose interior current is opposite to the initial current. The
standard method to calculate the decay of metastable state by quantum
tunneling is the bounce method in the Euclidean path integral. Once we
know the bounce solution of the Euclidean field equation satisfying
the correct boundary condition, the imaginary correction to the energy
density of the metastable state is ${\rm Im} E/V = Ae^{-B}$ where the
exponential suppressing factor $B$ is the difference between the
bounce action and the action for the metastable configuration and the
prefactor is naively the ratio of the path integral prefactors. In the
prefactor for the bounce solution, we take out the zero modes and take
half of the absolute value of negative eigenvalue. The decay rate per
unit volume is then $\Gamma = 2 \, {\rm Im} E/V$, where $V$ is the
volume factor.

Since our initial configuration is expected to decay due to the
nucleation of vortex rings and their expansion, it would be useful to
describe the vortex dynamics more explicitly. This is possible  in
the dual formulation of the scalar theory [11]. For a given
configuration of vortices, the $\psi(x)$ is multi-valued. The position
of the $a$-th vortex would be given by $q_a^\mu(\sigma^\alpha)$ with the
world sheet coordinates $\sigma^\alpha, \alpha=0,1$. For such a
configuration, there is a  conserved
vortex current $K^{\mu\nu}(x) \equiv \epsilon^{\mu\nu\rho\sigma}
\partial_\rho\partial_\sigma \psi$, which can be expressed as
$$ K^{\mu\nu} = 2\pi \sum_a \int d^2 \sigma \epsilon^{\alpha\beta}
\partial_\alpha q^\mu \partial_\beta q^\nu \delta^4(x^\rho -
q^\rho(\sigma)).
\eqno\eq  $$
The orientation of the world sheet coordinates fixes the orientation
of the vortex string, or the increasing  direction of the  phase
variable ${\bf \psi}$.

In the dual formulation,  the generating functional in the original
$f,\psi$ fields  is  rewritten in terms of the $f$, an
antisymmetric tensor field $C_{\mu\nu}$ and the vortex string position
$q^\mu_a(\tau, \sigma)$. The  on-shell relation between the original and
dual variables  is  $ f^2\partial^\mu \psi = {1\over 6}
\epsilon^{\mu\nu\rho\sigma} H_{\nu\rho\sigma}  $
where the field strength is  $H_{\mu\nu\rho} = \partial_\mu C_{\nu\rho} +
\partial_\nu C_{\rho\mu} + \partial_\rho C_{\mu\nu} $.
The dual Lagrangian is then given by
$$ {\cal L}_D = {1\over 2} (\partial_\mu f)^2 + {1\over 12
f^2}H_{\mu\nu\rho}^2 - U(f) + {1\over 2} C_{\mu\nu}K^{\mu\nu}.
\eqno\eq $$
{}From the Lagrangian (9), the canonical momentum density for $C_{ij}$
is $ H_{0ij} /f^2 $. Since $H_{012}/f^2$ for the homogeneous initial
configuration is $ j/u^2 $, the wavefunction for the initial
configuration is then proportional to $\exp\big\{ i\int d^3 x \, j
C_{12}/u^2 \bigr\} $. In the dual formulation, the uniform current
becomes the uniform ``electric'' field and exerts the ``electrostatic''
force on vortex strings, and so we expect vortex string loops to
nucleate.

The Euclidean path integral formalism is used here to calculate the
imaginary shift of the energy density, which leads to the decay rate.
We are calculating the  Euclidean path integral,
$$ <F |e^{-HT} |I > = \int [f^{-3}df dC_{\mu\nu}dq^\mu_a]\,  \Psi_F^* \,
e^{-S_E} \, \Psi_I
\eqno\eq $$
by the saddle point method.  The Euclidean action can be obtained by
the Wick rotation $t = -i \tau$ of the Minkowski action or by a direct
dual transformation of the Euclidean path integral of the complex
scalar field, and is given by
$$ S_{E} = \int d^4 x_E \left\{ {1\over 2} (\partial_\mu f)^2
+ {H_{\mu\nu\rho}^2 \over 12 f^2} + U(f)
-{i \over 2}  C_{\mu\nu} K^{\mu\nu} \right\}.
\eqno\eq $$
Note that the interaction between vortices and $C_{\mu\nu}$ is
topological and is not affected by the Wick rotation.
Since the initial and final configurations carry a nonzero current,
there is a boundary term in the exponent of the path integral (10),
$$ \Sigma = - i {j \over u^2}  \int d^3x \left\{  C_{12}(\tau_F, \vx)  -
 C_{12} (\tau_I,\vx) \right\}.
\eqno\eq $$
Since the $C_{\mu\nu}$ has the free boundary condition, the saddle
point of the path integral will be the the stationary field
configuration of the combined action, $ \tilde{S} = S_E + \Sigma $
which is also complex. {}From this combined action we get  a bounce
equation. The appropriate boundary conditions at $\tau=\pm \infty$ is
$f \rightarrow u $.

Since the Euclidean initial configuration of the uniform current along
the $z$ axis has the symmetry under three-dimensional rotations in the
$\tau,x,y$ coordinates, we look for the bounce solution of this $O(3)$
symmetry. With $r^2 \equiv \tau^2 + x^2 + y^2$, the solutions would
depend only on $r,z$.  Since the boundary term is purely imaginary,
we see that the $C_{\mu\nu}$ field for the bounce solution should be
pure imaginary to satisfy the boundary condition. We also choose the
ansatz so that the $C_{\mu\nu}$ field has the spherical symmetry,
$$ {\bf C} \equiv {1\over 2} C_{\mu\nu} dx^\mu dx^\nu = i C(r,z) \sin
\theta d\theta d\varphi.
\eqno\eq $$
The field strength ${\bf H} = d {\bf C} $ would be
$$ \eqalign{ {\bf H} = {1\over 6} H_{\mu\nu\rho} dx^\mu dx^\nu dx^\rho
= i \partial_r C \sin \theta dr d\theta d\varphi + i \partial_z  C
\sin \theta dz d\theta d\varphi \cr} .
\eqno\eq $$
The boundary term (12) leads to the boundary condition $C(r,z)
\rightarrow jr^3/3$ for large $r$. For the field configuration to be
smooth at the line $r=0$, $C(r,z)$ should vanish when $r=0$.

{}From the translation symmetry along the $z$ axis, we choose
that the string lies on the $z=0$ plane and  a
two-dimensional sphere of a given radius $R$ in the
$\tau,x,y$ space.  The string current is invariant of the
choice of the string internal coordinate.  We choose the
string coordinate so that  $ q^\mu(\sigma^\alpha) = (R \sin \sigma^0
\cos \sigma^1 ,R \sin \sigma^0 \sin \sigma^1, R \cos \sigma^0 , 0)
$. The delta function $\delta^4(x^\rho -q^\rho) $ becomes
$\delta(r-R)\delta(\theta -\sigma_0 )\delta(\varphi-\sigma_1)
\delta(z)/R^2\sin\theta$.  The string tensor current (9) becomes
$$  K^{\mu\nu}(x)
 =  2\pi  ( \hat{\theta}^\mu
\hat{\varphi}^\nu - \hat{\theta}^\nu \hat{\varphi}^\mu )   \delta( r
- R) \delta(z),
\eqno\eq $$
where $\theta^\mu = (\hat{\theta},0), \varphi^\mu =
(\hat{\varphi},0)$.  If it was the nucleation of a vortex with
opposite vorticity along the $z$ direction, there would have been a
negative sign on the above vortex current.  For the spherical
symmetric solutions, the bounce field equation from the combined
action is given as
$$ \eqalign{ &\ {1\over r^2} \partial_r ( r^2
\partial_r f   ) + \partial^2_z f
 - {1\over r^4 f^3} \left\{ (\partial_r C)^2
 + (\partial_z C )^2 \right\}
- U'(f) =  0 \cr
&\ r^2 \partial_r\left( {1\over r^2   f^2} \partial_r C \right) +
\partial_z  \left( {1\over f^2}
\partial_z C   \right) = 2\pi R^2 \delta(r-R) \delta (z).       \cr}
\eqno\eq $$
There cannot be a regular solution satisfying the boundary condition
for arbitrary $R$. The reason is that we expect only one bounce solution,
resulting in a unique radius. Except for this value of $R$, we expect that
the solution satisfying the boundary condition at the infinity will
become singular along  the line $r=0$.  The combined action for this
configuration can be put as
$$ S_E + \Sigma  = 4\pi  \int  dr r^2 dz  \left\{
{1\over 2} (\partial_r f)^2  + {1\over 2}  (\partial_z  f)^2
+ {1\over 2  r^4 f^2} [(\partial_r  C)^2 + (\partial_z C)^2] + U(f)
\right\}.
\eqno\eq  $$

The bounce solution $f_b(\tau,\vec{x}), C_b(\tau,\vec{x})$ of Eq.(16)
can be analytically continued to lead to the solution of the Minkowski
time.  Let us first express the antisymmetric tensor field in the
Cartesian coordinate, noting $\tan\theta = \rho / \tau$ with $\rho =
\sqrt{ x^2 + y^2}$, the two-form (13) becomes $ {\bf C}_b = i C_b(r,z)
(\rho^2  d\tau  + \rho\tau / r^3 d\rho ) d\varphi/r^3 $.
In  the  $\rho>t$ region, the solution in Minkowski time is
$$ \eqalign{ &\ f(t, \vec{x}) = f_b(it, \vec{x}) \cr
&\ {\bf C}(t, \vec{x}) =    C_b(\sqrt{\rho^2-t^2},z) \biggl( {t^2
\over r^3} d\rho d\varphi - {\rho t \over r^3} dt d\varphi \biggr) \cr},
\eqno\eq $$
which leads to nonzero $C_{\rho\varphi}, C_{t\varphi}$. By taking the
exterior derivative of ${\bf C}$, we can get the field strength.  We
can further analytically continue the above solution to the $\rho< t$ region.
{}From the Euclidean two-sphere covered by the vortex string, $ x^2 +
y^2 + \tau^2 = R_b^2$, we get the radius for the vortex string on the
$x-y$ plane in Minkowski time as $ R(t) = \sqrt{ R_b^2 + t^2} $. The
size of the string then increases with a constant acceleration.

For a given initial current, it is not easy to find the bounce
solution of Eq. (16) and get the tunneling rate. When the initial
current is much smaller than the critical current, the size of the
two-sphere would be much larger than the Compton wave-length of the
Higgs particle and the $f$ would be very close to $u$ in this case
away from the two-sphere.  The bounce solution can then be well
approximated by the solution of Eq.(16) with $f=u$ outside the core
region. Later we will show that this approximation is reasonably good.

One conventional approach to this problem is to find a one-parameter
$\lambda$ family of field configurations along the tunneling path
and to calculate the effective energy for these configurations,
$E(\lambda) = K(\lambda) \dot{\lambda}^2 + V(\lambda)$. We can then
treat the problem by the one-dimensional WKB method. The tunneling
path in the configuration space is then described by these
one-parameter field configurations.  In our case, we can have an
effective string action coupled to the antisymmetric tensor fields.
{}From this, we can get the energy functional for a vortex string ring
as a function of its radius.  However, the self-interaction term is
complicated to treat.

We follow another approach [2], which considers the one-parameter
family of field configurations $\phi(\lambda,x)$ in Euclidean time.
Each field configuration is a solution of the Euclidean field
equation, which might be mildly singular.  Here, we imagine  exploring
the configuration space with these one-parameter stationary field
configurations.  The Euclidean action $S(\lambda)$ of the
configurations in this family is a finite function of this parameter
and we find the stationary point among this family with respect to
this parameter.  The configuration at the stationary parameter would
be the stationary configuration we looked for. The radius $R$ of the
two-sphere is the parameter. This approach is more profitable when
there is a long-range force in the system as it treats the space and
time components of the force simultaneously.

When  $f=u$, the bounce equation (16) becomes
$$ r^2 \partial_r \left( {1\over r^2 }\partial_r C(r,z)\right)
+ \partial_z^2 C(r,z) =  2\pi u^2 R^2 \delta(r-R)\delta(z).
\eqno\eq $$
Since the equation is linear, we can require $C$ to satisfy the
boundary condition at infinity and to be zero on the line $r=0$.  We
approach this by solving the homogeneous equation by  the fourier
transformation in the $z$ coordinate,
$$ C(r,z) = {1\over 3} jr^3 +  \int dp e^{ipz} \left(  a_p (  pr - 1)
e^{pr} + b_p(pr +1)e^{-pr}  \right).
\eqno\eq $$
We impose the boundary conditions and match the homogeneous solutions
at $r=R$.  Our  solution  of Eq.(19)  is then the sum
of the background part
$$ C_{\rm back}(r)  = {1\over 3} jr^3
\eqno\eq $$
and the dynamical  part
$$ C_{\rm dyna}  =  -  {u^2 \over 4}\left\{
(R^2 + r^2 + z^2) \ln \left[ {(R+r)^2 +
z^2  \over (R-r)^2 + z^2 } \right] - 4Rr \right\} . \eqno\eq $$
There are several regions where we are interested in the detail
behavior of $C_{\rm dyna}$. Near the two-sphere,  $(R-r)^2 + z^2 << R^2$
and
$$ C_{\rm dyna}  \approx  - { u^2 R^2\over 2} \left\{ ln
\left[ { (R-r)^2 + z^2) \over 4R^2}\right]  + 2  \right\}.
\eqno\eq $$
We can read off the delta function  in Eq. (19) from this expression.
Far away from the sphere,  $r^2 + z^2 >>R^2$ and  $ C_{\rm dyna}
\approx - 4u^2  r^3  R^3 / 3(r^2+z^2)^2  $.
For near the center of the sphere,  $r^2+z^2<< R^2$ and
$ C_{\rm dyna}  \approx   -4u^2r^3 / 3R $.

Now we can ask how good our approximation $f=u$ is.
Far away from the two sphere, the $C(r,z)$ approaches
$C_{\rm back}$ and so we can see that $f\approx u$ becomes the
solution of the first of Eq.(16) due to Eq.(2).
{}From Eq.(16) we can get the equation satisfied by the
correction $\delta f = f-u$ due to the $C_{dyna}$.
Away from the two sphere, we can see $\delta f$ is approximated by
$$ \delta f = {2j \partial_r C_{dyna} \over r^2 u^3 U''(u)}
\eqno\eq
$$
to the first order in $C_{dyna}$. We studied the asymptotic behavior
of $C_{dyna}$ in various regions in the previous paragraph. We
estimate $ \delta f= -8j / (uU''(u)R )$ near the origin and $\delta f$
vanishes quickly for  $r^2 + z^2 >> R^2$. Thus the correction is
quite small if $R$ is much larger than the mass parameter.  We will
later argue that the correction to the action due to $\delta f$ is
also small compare with that from $C(r,z)$, making the whole
approximation consistent.

As  the $f=u$ is a good approximation except near the two-sphere,
our configuration $f=u, C = C_{\rm back} + C_{\rm dyna}$ is an
approximately stationary configuration of the action away from the
two-sphere. Because  our approximate solution does not have any
singularity at $r=0$, we expect the singularity at $r=0$ of the
solution of Eq.(16) for arbitrary $R>>m^{-1}$ will be  mild. Our
approximate solution is singular near the two-sphere of distance
less than $m^{-1}$.  We now have a one-parameter family of the field
configuration whose is approximately stationary except  along the
parameter $R$.

Rather then calculating the action (17), which diverges at the
infinity, we calculate the exponential suppression factor $B$, which is
the difference between the bounce action (17) and the action for the
initial configuration. We write the exponential factor $B$ as a sum
$$ B(R) =   \int d^4x \bigg\{ {1\over
12u^2} \bar{H}_{\mu\nu\rho}^2 \biggr\}
  - (S_E + \Sigma)({\rm background}),
\eqno\eq $$
where $\bar{H}$ is the field strength for our approximate solution
given by Eqs.(21) and (22).

The action $B$ is the difference between the action for the
approximate bounce solution and the background action.  Since the
$C(x)$ is a sum of the background field $C_{\rm back}$ and $C_{\rm
dyna}$, we get the $B$ as a sum
$$ \eqalign{ B(R)= &\  {4\pi } \int {dzdr r^2 \over u^2r^4} \biggl\{
\partial_r C_{\rm back} \partial_r C_{\rm dyna} + \partial_z C_{\rm dyna}
\partial_z C_{\rm dyna} \biggr\} \cr
&\ + {4\pi} \int {dz dr r^2 \over 2u^2r^4} \biggl\{    (\partial_r
C_{\rm dyna})^2  + (\partial_z C_{\rm dyna})^2 \biggr\}. \cr}
\eqno\eq $$
In the integration, we introduce  the short distance cutoff $\epsilon$ of
order $m^{-1}$ near the two-sphere. This is because our approximate
configuration becomes singular there. The first integral of Eq.(26)
accounts for the interaction between the background `electric' field and
the vortex string loop. The second integral is the self-interacting
part of the vortex string loop.  The integral can be performed by
introducing two cutoffs $ (r-R)^2 + z^2 \ge \epsilon^2$ and $r^2 + z^2
\le L^2$, by using the integration by parts and by using the
asymptotic behaviors of the $C_{{\rm int}}(x)$ fields in various
regions. The short distance cutoff $\epsilon$ near the two-sphere is
of order $ m^{-1}$ . Putting these together, we get
$$ B(R)= 4\pi^2 u^2 R^2  \ln ( c mR)  - 2\pi^2 j R^3,
\eqno\eq $$
where $c$ is a constant of order one.

Let us now consider what are the correction to the above equation.
The solution of Eq.(16) is not singular near the two sphere and the
correction to the factor $B$ from the region inside the cutoff near
the two sphere would be of order $m^2 R^2$ from the dimensional
analysis.  Since $\delta f \approx 1/R$, one can see the correction to
$B$ due to $\delta f$ would be again of order $m^2 R^2$. Both
corrections change the coefficient $c$ by order 1, making Eq.(27)
reasonable.

Now we can find the stationary point of the exponential factor $B$
with respect to $R$ and get the bounce radius $R_b>> m^{-1}$ as
$$ R_b \approx {4u^2 \over 3j}  \ln \left( {4 \sqrt{e} cmu^2
\over 3j} \right)
\eqno\eq  $$
and the exponential suppression factor  at this radius becomes
$$ B_b \approx   {64 \pi^2 u^6 \over 27 j^2} \bigg\{ \ln \left(
{4\sqrt{e}cm u^2 \over 3j}\right) \biggr\}^3.
\eqno\eq $$
By a simple scaling argument $ f/v, C/v^2, r\sqrt{\lambda}v$, we can
show that the classical action for the bounce is of the form $ B_b =
{1\over \lambda} F( {j_c \over j}) $ with an unknown function $F$.
Eq.(29) provides an approximate form for the function $F$.  The radius
(28) of the bounce solution, which is the critical radius, takes a
form $ mR \sim (j_c/j) \ln(j_c/j) $ and the bounce action (29) is of
order $(j_c^2/\lambda j^2) (\ln(j_c/j))^3$. Thus our semiclassical
approximation is good if $j<< j_c$ and $\lambda <<1$.

Let us now consider briefly the one-loop correction to our bounce
solution.  We can expand the fluctuations around the bounce solution
in the spherical harmonics. The antisymmetric tensor field should be
expanded in the tensor spherical harmonics. There will be four zero
modes for the translation.  There will be one negative mode which
corresponds to the change of the radius for two-sphere. There will
also be gauge fixing terms for the antisymmetric tensor field. Since
the original theory is renormalizable, we expect the dual formulation
would somehow be renormalizable, even though the dual theory is a
strong coupling theory.  In principle, the ultraviolate divergence
should cancel that from the background.  {}From the standard technique
[9,2], it is not hard to see that the normalization factor for each
zero mode is of order $1/R$.  The one-loop correction would be then
$$ K =  {a \over R^4}
\eqno\eq $$
with a dimensionless number  $a$. The precise factor does not matter
in our case because there is an uncertainty of order $R$ in the $ B$
factor. By combining the results, the vortex string loop creation rate
per unit volume becomes $ \Gamma_V = 2K e^{-B_b}$.

In addition, we notice that  a  similar question can be addressed to
the Maxwell-Higgs systems with the Lagrangian
$$ {\cal L} = -{1\over 4e^2}F_{\mu\nu}^2  + {1\over 2}( \partial_\mu f)^2
+ {1\over 2} f^2 (\partial_\mu \theta + A_\mu)^2 - U(f) - J^\mu_{ext} A_\mu.
\eqno\eq $$
Without the external current, the system cannot be homogeneous in
the presence of a uniform matter current due to the nonzero magnetic
field.  The initial configuration is a homogeneous configuration $f=u,
A^z= j/u^2$ with an external current $J^\mu_{ext} = (0,0,0,j)$. The
current carried by the Higgs field cancels the background electric
current, resulting in the zero total current.  For this homogeneous
solution to be static the $u$ satisfies Eq.(2).  {}From the analysis
of the small fluctuation around this configuration, one can show that
this configuration is also classically stable if Eq.(4) is correct. As
we are in the Higgs phase, there is no massless mode and instability
can occur if the mass becomes negative.  With the specific potential
given in Eq.(5), such a homogeneous configuration can exist only if
the external current is smaller than the critical current (7). The
classically stable homogeneous configurations can again decay quantum
mechanically by the nucleation of local vortex string loops. The
analysis of such tunneling process would be somewhat similar to the
global vortex nucleations in the previous discussions. We can use the
dual formulation in the second paper of Ref.[11]. There is again an
$O(3)$ symmetric bounce solution.  The field configuration away from
the two-sphere will fall exponentially to the initial configuration.
Hence, there is no logarithmic term in the bounce action. The analog
of Eq.(28) is $B(R) = 4\pi \mu R^2- 2\pi^2j R^3$, where $\mu$ is the
string tension in the small current limit.  The bounce radius becomes
$R_b = 4\mu / 3\pi j $ and the exponential suppressing factor becomes
$ B_b = 64\mu^3/(27 \pi j^2)$.

We have discussed the vortex loop nucleation in a uniform background
current. We have analyzed the nucleation rate per unit volume and the
size of the vortex loop at the moment of the nucleation. We argued
that its size grows with a constant acceleration. We then extend our
discussion to a Maxwell-Higgs theory with a uniform external current.
We can approach our problem by an effective string action which
contains the Nambu action and the interaction term between the
background `electric' field and strings. This would lead to a similar
answer obtained previously if we take into account the string
self-interaction.

The two-fluid model for the superfluid works fine in describing the
vortex loop nucleation in finite temperature [6]. Vortex loops
nucleate thermally when the relative speed between two fluids is
nonzero. The vortex nucleation in that case does not allow the naive
path integral treatment as we do not know how to treat the normal
fluid in the path integral. However, we can imagine a somewhat similar
case where  two superfluids exist at zero temperature, whose currents
are both timelike but not identical. If we imagine a nontrivial
coupling between them, vortices might nucleate. This process can be
approached by the path integral. The characteristics of vortex loop
nucleation will depend on the charge coupling. Since the current is
timelike, there may be some surprise here.

Finally, we can ask what is the final configuration of the string
nucleations. (If we imagine a finite volume, string loops will be
created and annihilate each other and disappear from the system. The
initial current will be continuously reduced.)  Thus we expect the
final configuration to be some sort of superfluid turbulence. In the
theory with the local symmetry, the initial current of the Higgs field
will decay away but the external current is frozen. We end up with a
curious situation where there is the external current but no
compensating current by the Higgs field.  This configuration if static
in time seems nonsensical because the static solution of the Maxwell
equation has the the magnetic field growing linearly in space and so
the energy density grows quadratically.  The final configuration can
be obtained by considering a similar system in two dimensional
spacetime where the initial current decays by the bounce solution with
vortex and antivortex lying on a Euclidean time axis. The initial
Higgs current again decays away. The electric field grows linearly in
time. However, one can show that the energy is conserved because there
is no lower bound on the energy functional due to the external
current. In short, there is no ground state in the local gauge theory
when there is the external current.

\vskip 0.4in

\centerline{\bf Acknowledgments}

The work by H.K. is supported by National Science Council, Taiwan
(Grant No. NSC84-2112-M-001-022). The work by K.L.  is supported in
part by the NSF Presidential Young Investigators program, the Alfred
P. Sloan Foundation.

\vskip 1.0in

\noindent {\bf  note added:}

After this work appeared in a preprint form, we are informed by Dr.
J.-M. Duan about his related works on the vortex nucleations in the
charge density wave systems [12].
\endpage

\refout

\end